\documentstyle[epsf,amsmath]{article}

\newcommand{\lh}[1]{\slash \! \! \!  #1}

\begin{document}
\setcounter{page}{0}
\mbox{ }
\rightline{UCT-TP-263/04}\\
\rightline{Revised January 2005}\\
\vspace{3.5cm}
\begin{center}
{\LARGE \bf Electromagnetic nucleon form factors from\\
 QCD sum rules}\footnote{Work supported in part by Fondecyt
grants No.1010976 and 7010976}\\
\vspace{.5cm}
{\Large H. Castillo $^{(a),(b)}$ , C.A. Dominguez $^{(c)}$, 
M. Loewe $^{(a)}$}\\
\vspace{.4cm}
\end{center}
(a) Facultad de F\'{\i}sica, Pontificia Universidad Cat\'{o}lica  de Chile, 
Casilla 306, Santiago 22, Chile.\\
(b) Departamento  de 
Ciencias,  Pontificia Universidad Cat\'{o}lica del Per\'{u}, Apartado 1761,
 Lima, Per\'{u}.\\
(c) Institute of Theoretical Physics \& Astrophysics, 
University of Cape 
Town, Rondebosch 7701, South Africa.\\
\vspace{.5cm}
\begin{abstract}
\noindent
The electromagnetic form factors of the nucleon,
in the space-like region, are determined from 
three-point function Finite Energy QCD Sum Rules.
The QCD calculation is performed to leading order
in perturbation theory in the chiral limit, and to 
leading order in the non-perturbative power corrections.
The results for
the Dirac form factor, $F_1(q^2)$, are
in  very good agreement with  data for both the proton
and the neutron, in
the currently accessible experimental region 
of momentum transfers. This is not the case, though, for the
Pauli form factor $F_2(q^2)$, which has a soft $q^2$-dependence
proportional to the quark condensate $<0|\bar{q}q|0>$.

\end{abstract}
\newpage
\setlength{\baselineskip}{1.5\baselineskip}
\noindent

The  electromagnetic form factors of the nucleon 
have been studied in perturbative QCD (PQCD), together with QCD sum rule
estimates of the nucleon wave functions \cite{REV1}.  Comparison with data is 
difficult due to the extreme asymptotic
nature of these theoretical results. In fact, the onset of PQCD in exclusive reactions
does not appear to be as precocious as in  inclusive processes. 
In addition, these wave functions are affected by 
some unavoidable model dependency. In any case,
the Dirac form factor $F_1(Q^2)$ does exhibit the expected leading asymptotic
$1/Q^4$ behaviour. However, the Pauli form factor $F_2(Q^2)$ turns out to be
of higher twist, and therefore not accessible in the standard PQCD approach.
At current experimental space-like momentum transfers, the results from the
standard hard-scattering approach for $F_1(Q^2)$ do not compare favourably
with the data. On the other hand, some recent light-cone QCD sum rule
determinations appear to improve the agreement with data from within a
factor 5-6 to within a factor of two \cite{BRAUN}. The source of this persistent
disagreement does not seem easy to identify.
In view of this,  it is desirable to attempt a  QCD sum rule determination in a 
region of experimentally accessible momentum transfers, and without  any
reference to the concept of a wave function. In addition, one should employ 
sum rules of a type which would provide a clear insight into the source(s) of
potential disagreement with experiment. This can be achieved e.g. by using
Finite Energy Sum Rules (FESR). In fact, in this framework the power
corrections involving the vacuum condensates decouple to leading
order in PQCD. In other words, power corrections of different dimensionality
contribute to different FESR. \\

In this note we determine the Dirac and Pauli electromagnetic nucleon form
factors, in a wide range of (space-like) momentum transfers, in the framework of 
three-point function QCD-FESR of leading dimensionality. As is well known by now,
this technique  is based on the Operator Product Expansion (OPE) of current 
correlators at  short distances, and on the notion of quark-hadron duality
\cite{QCDSR}. 
Analyticity and dispersion relations connect  the QCD information in the 
OPE to hadronic parameters entering the corresponding spectral functions. 
We compute the QCD correlator  to leading order in perturbative QCD  in the 
chiral limit ($m_u=m_d=0$), and include the leading order non-perturbative 
power corrections proportional to the quark-condensate and the four-quark
condensate (with no gluon exchange).
We begin by considering the following three-point function (see Fig. 1)\\
\begin{equation}
\Pi_\mu (p^2,p'^{2},Q^2) =  i^2 \!\int \!\!\mbox{d}^ 4 x \!\!\int
 \!\!\mbox{d}^4 y \; e
^{i (p' \cdot x - q  \cdot y)}\, \langle 0 \left| T \{ \eta_N (x)
 J_\mu^{EM}(y)\bar
 \eta_N  
(0) \}
\right| 0\rangle \; , 
\end{equation}

where $Q^2\equiv - q^2 = -(p' - p)^2\geq 0$ is fixed, and
\begin{equation}
\eta_N(x)=\varepsilon_{abc}\left[u^a(x)(C\gamma_\alpha)u^b(x)\right]
(\gamma^5\gamma^\alpha d^c(x)) \label{ec2}
\end{equation}
is an interpolating current with nucleon (proton) quantum numbers; the 
neutron case $u \leftrightarrow  d$ will be discussed at the end.
In Eq.(1),  $J^{\mu}_{EM}$ is the electromagnetic current\\
\begin{equation}
J_{EM}^\mu (y)=\frac 2 3 \bar u(y)\gamma^\mu u(y)-\frac 1 3 \bar 
d(y)\gamma^\mu d(y) \; .
\end{equation}
The current Eq.(2) couples to a nucleon of momentum $p$ and polarization $s$ 
according to
\begin{equation}
\langle 0\left| \eta_{N}(0) \right|N(p,s)\rangle = \lambda_N u(p,s) ,
\end{equation}
where $u(p,s)$ is the nucleon spinor, and $\lambda_N$, the current-nucleon 
coupling, is a phenomenological parameter a-priori unknown. This parameter 
can be estimated, e.g. using QCD sum rules for a two-point function 
involving the currents $\eta_N$ \cite{QCDSR}-\cite{MN}. In this case one can determine 
the nucleon mass, as well as the coupling $\lambda_N$.\\
Concentrating first on the hadronic sector, and inserting a one-particle 
nucleon state in the three-point function (1) brings out the nucleon form 
factors $F_1(q^2)$, and $F_2(q^2)$,
defined as
\begin{equation}
\langle N(k_1\,, s_1)\left|  J_\mu^{EM}(0) \right| N(k_2\,,s_2)\rangle=
\bar u_N (k_1,s_1)
\left[ F_1 (q^2) \gamma _\mu + \frac{i \kappa}{2M_N}
F_2(q^2)\sigma_{\mu \nu} q^\nu \right]  
u_N(k_2,s_2) \; ,
\end{equation}
where $q^2= (k_2-k_1)^2$, and $\kappa$ is the anomalous magnetic moment
in units of nuclear magnetons ($\kappa_p = 1.79$ for  the proton, and
$\kappa_n = - 1.91$  for the neutron). The form factors 
$F_{1,2}(q^2)$ are related to the electric and magnetic (Sachs) form factors 
$G_E(q^2)$, and $G_M(q^2)$, measured in elastic electron-proton scattering 
experiments, according to
\begin{eqnarray}
G_E(q^2) &\equiv& F_1(q^2)+\frac{\kappa q^2}{(2m)^2} F_2(q^2)\; ,
\\ [.2cm]
G_M(q^2) &\equiv& F_1(q^2)+ \kappa F_2(q^2) \; ,
\end{eqnarray}
where $G_E^p(0) = 1$, $G_M^p(0) = 1 + \kappa_p$ for the proton, and
$G_E^n(0) = 0$,  $G_M^n(0) = \kappa_n$ for the neutron.
Next, the hadronic spectral function is obtained after inserting a complete 
set of  nucleonic states in (1), and computing the double discontinuity in the 
complex $p^2 \equiv s$, ${p'}^2 \equiv s'$
plane. For $s, s' < 2.1\, \mbox{GeV}^2$, i.e. below the Roper resonance, one 
can safely approximate the hadronic spectral function by the single-particle 
nucleon pole, followed by a continuum with thresholds $s_0$ and $s'_0$ 
($s_0, s'_0 > M_N^2$). This hadronic continuum is expected to coincide 
numerically with the perturbative QCD (PQCD) spectral function (local
duality). 
This procedure is standard in QCD sum rule applications, and leads to
\begin{equation}
\begin{aligned}
\mbox{Im} \Pi_\mu(s,s',Q^2)\Big|_{HAD}&=\pi^2\,\lambda_N^2\, \delta
(s-M_N^2)
\delta(s' - M_N^2) \\
 & \times \Bigg \{ F_1(q^2) \left[\lh p' \gamma_\mu \lh p+M_N (\lh p' \gamma_\mu+
\gamma_\mu \lh p)+ M_N^2 \gamma_\mu \right]\\ &+\frac{i \kappa}{2
M_N}F_2(q^2)\left[\lh p' \sigma_{\mu\nu}\lh p+M_N(\lh p'
\sigma_{\mu\nu}+\sigma_{\mu\nu}\lh p)+M_N^2 \sigma_{\mu\nu} 
\right] q^\nu
\Bigg \} \Theta\left(s_0-s\right)  \\
&+\mbox{Im} \Pi_{\mu}(s,s',Q^2)\Big|_{PQCD}\Theta\left( s-s_0\right)
\; ,
\end{aligned} 
\end{equation}
where we have set $s_0=s'_0$ for simplicity.\\

Turning to the QCD sector, the three-point function (1) to leading order in 
perturbative
QCD, and in the chiral limit, is given by\\
\begin{equation}
\begin{aligned}
\Pi^\mu(p^2,{p'}^2,Q^2)= 16 \int \mbox{d}^ 4 x \; \int \mbox{d}^ 4 y 
\; e^{i(p' \cdot x-q
\cdot y)}\; \mbox{Tr}\left[\int\frac {\mbox{d}^4 {k_1}} {(2\pi)^4} 
\frac {\lh {k_1}}{k_1^2}
e^{-i k_1 \cdot (x-y)}\gamma^\mu \;  \right. \\ \left. \times \int\frac 
{\mbox{d}^4 
{k_2}} {(2\pi)^4} \frac {\lh {k_2}}{k_2^2} e^{-i k_2 \cdot y} \gamma_\nu 
\int\frac {\mbox{d}^4 {k_3}} {(2\pi)^4} \frac {\lh {k_3}}{k_3^2}
e^{+i k_3 \cdot x} \gamma_\alpha\right]\left(\gamma^5\gamma^\alpha 
\int\frac {\mbox{d}^4 {k_4}} {(2\pi)^4} \frac {\lh {k_4}}{k_4^2}
e^{-i k_4 \cdot x}
\gamma^\nu\gamma^5\right)\\
+ \;4 \int\mbox{d}^4 x\int \mbox{d}^4 y \;e^{i( p' \cdot x-q \cdot y)} \;
\mbox{Tr}\left[
\int\frac {\mbox{d}^4 {k_4}} {(2\pi)^4} \frac {\lh {k_4}}{k_4^2}
e^{-i k_4 \cdot x} \gamma_\nu 
\int\frac {\mbox{d}^4 {k_3}} {(2\pi)^4} \frac {\lh {k_3}}{k_3^2}
e^{+i k_3 \cdot x} \gamma_\alpha\right]\\
\times \left(\gamma^\alpha 
\int\frac {\mbox{d}^4 {k_1}} {(2\pi)^4} \frac {\lh {k_1}}{k_1^2}
e^{- i k_1 \cdot (x-y)}  \gamma^\mu 
\int\frac {\mbox{d}^4 {k_2}} {(2\pi)^4} \frac {\lh {k_2}}{k_2^2}
e^{-i k_2 \cdot y} \gamma^\nu\right).
\end{aligned}
\end{equation}

After computing the traces and performing the momentum space integrations, 
Eq.(9) involves several Lorentz structures analogous to those entering
the hadronic spectral function  Eq. (8).
Before invoking duality one needs to choose a particular Lorentz structure 
present in both (8) and (9). A convenient choice turns out to be 
$\lh{p'} \gamma_\mu  \lh{p}$, which
allows to project $F_1(q^2)$, as this structure does not appear multiplying 
$F_2(q^2)$ in Eq.(8).  An additional advantage of this choice is that the 
quark condensate  contribution, to be discussed later, does not involve the
structure  $\lh{p'} \gamma_\mu \lh{p}$, on account
of vanishing traces. There is, though, a non-perturbative term involving this
structure and proportional to the four-quark condensate. However, eventually
this term will  not contribute to the FESR as 
its double discontinuity vanishes.  Hence, $F_1(q^2)$ will only be dual  
to the PQCD  expression. It must be pointed  out that the PQCD spectral 
function contains the structure 
$\lh{p'} \gamma_\mu \lh{p}$ explicitly, as well as implicitly,  i.e. 
there are terms proportional to this structure which are generated only 
once the momentum-space integration is performed.\\
After a very lengthy calculation, the imaginary part of Eq.(9) is given by\\

\begin{equation}
\mbox{Im} \Pi^\mu(s,s',Q^2) =\left[\frac 4 {(2\pi)^8}(3
\Omega_1+ 4 \Omega_2-\Omega_3)\right]\left(\lh p' \gamma^\mu
\lh p\right)+\ldots \; ,
\end{equation}
where\\

\begin{equation}
\Omega_1=\frac{{\pi }^6} 2 \,\left( Q^2 + s - s' -  \frac{Q^4 + 
2\,Q^2\,s +
s^2 - 2\,s\,s' -  {s'}^2}{{\sqrt{Q^4 +
          {\left( s - s' \right) }^2 +
          2\,Q^2\,\left( s + s' \right) }}} \right)\; ,
\end{equation}

\begin{eqnarray}
\Omega_2&=&{\pi}^6 \left\{\frac{\left( 2\,Q^2 + 3\,s - 3\,s' \right) }{3}
-\frac{\left[( Q^2 + s)^3( 2\,Q^2 + 3s) + 
       3\,\left( Q^6 - 5Q^2s^2 - 4s^3 \right)s'\right]}
 {3{\left[ Q^4 + {( s - s')}^2 + 
         2\,Q^2( s + s')  \right] }^{\frac{3}{2}}} \right.\nonumber  \\[.2cm]
 &+& \left. \frac{ \left[ (3Q^2 - 4s) ( Q^2 + 3s){s'}^2 + 7Q^2{s'}^3 + 
       3{s'}^4 \right]} {3\left[ Q^4 + {( s - s')}^2 + 
         2\,Q^2( s + s')  \right] ^{\frac{3}{2}}}\right\} \; ,  \\
         [.9cm]
\Omega_3 &=&  \frac{-\left\{ \pi ^6\,\left[ 23\,Q^2 + 18\,\left( -s + s'
\right) \right]
\right\} }{72} 
+\frac{\pi^6}{72\,\left[ Q^4 + \left( s - s' \right) ^2 + 2\,Q^2\,
\left( s + s' \right)  \right] ^
\frac{5}{2}} \nonumber \\[.2cm]
&\times&  \left[ \left(23\,Q^2 - 18\,s \right) \,\left( Q^2 +
 s \right)^5  
+ \left( Q^2 + s \right) ^3\,\left( 133\,Q^4 - 169\,Q^2\,s + 108\,s^2
 \right) \,s' \right. \nonumber \\ 
&+& \left. 2\,\left( 160\,Q^8 + 6\,Q^6\,s + 3\,Q^4\,s^2 + 40\,Q^2\,s^3 - 117\,
s^4 \right)
\,{s'}^2  \right. \nonumber\\   
&+& \left. 2\,\left( 205\,Q^6 - 61\,Q^4\,s - 122\,Q^2\,s^2 + 108\,s^3 \right)
 \,{s'}^3  \right. \nonumber \\ 
&+& \left.  \left( 295\,Q^4 - 37\,Q^2\,s - 54\,s^2 \right) \,{s'}^4 +
 \left( 113\,Q^2 - 36\,s \right) \,{s'}^5 + 18\,{s'}^6  \right] \;.
\end{eqnarray}

Equation (11) corresponds to the terms containing  $\lh{p'} \gamma_\mu 
\lh{p}$ explicitly, and Eqs.(12)-(13) to the implicit case. The spectral 
function (10)  contains
additional terms proportional to other (independent) Lorentz structures, 
which are not written above. Collecting all three terms in (10) leads to
\begin{eqnarray}
\mbox{Im} \Pi^\mu(s,s',Q^2)&=&
\frac{323\,Q^2 + 378\,\left( s - s' \right) }{4608\,{\pi }^2} \;
+ \frac 1 {4608\,\pi ^2\,
\left[ Q^4 + ( s - s') ^2 + 2\,Q^2( s + s')\right]^{\frac{5}{2}}}
\nonumber \\ [.2cm]
&\times&
\left[ -323\,Q^{12} - Q^{10}( 1993\,s + 1237\,s')  
- 10\,Q^8( 512\,s^2 + 323\,s\,s' + 134\,{s'}^2)\right. \nonumber \\
&+&\left.  Q^6 ( -7010\,s^3 + 1188\,s\,{s'}^2 + 550\,{s'}^3 )
\right. \nonumber\\
&+& \left.  Q^4( -5395s^4 + 7010s^3s' + 2610s^2{s'}^2 + 3146s{s'}^3 
+2165{s'}^4) \right. \nonumber \\
&-& \left. Q^2\,( s - s') ^2 (2213\,s^3 - 2859\,s^2\,s' - 
3099\,s\,{s'}^2 - 1567\,{s'}^3) \right. \nonumber \\
&-&\left. 378\,( s - s')^4 (s^2 - 2\,s\,s' - {s'}^2 )\right]
\;\lh p'\gamma^\mu \lh p+\ldots
\end{eqnarray}

The next step is to invoke  (global) quark-hadron duality,  according to 
which the area under the hadronic spectral function equals the  area under 
the  corresponding QCD
spectral function. The integrals in the complex energy plane may involve any 
analytic integration kernel; this leads to different kinds of QCD sum rules, 
e.g. Laplace (negative exponential kernel), Finite Energy Sum Rules (FESR) 
(power kernel), etc. We choose the latter, as they have the advantage of 
being organized according to dimensionality (to leading order in gluonic 
corrections to the vacuum condensates). In this case the FESR of leading dimensionality is
\begin{equation}
\int_0^{s_0}\mbox{d} s
\int_0^{s_0-s} \mbox{d} {s'}\;\mbox{Im}\Pi(s,s',Q^2)\mid_{HAD}=
 \int_0^{s_0}
\mbox{d} s
\int_0^{s_0-s} \mbox{d} {s'}\; \mbox{Im}\Pi(s,s',Q^2)\mid_{QCD} 
\; .
\end{equation}
The integration region, shown in Fig. 2, has been chosen as a triangle; the 
main contribution being that of region I, and the area included from regions
II and III tends  to compensate
the excluded regions. Other choices, e.g. rectangular regions, lead to 
similar final results, as discussed in \cite{IOFFE81}-\cite{DLR}. After 
performing the integrations, one finally obtains
\begin{eqnarray}
F_1(Q^2)&=&\frac{2\,{s_0}\,\left( 96\,Q^6 + 297\,Q^4\,{s_0} + 158\,Q^2\,
{{s_0}}^2 -
112\,{{s_0}}^3 \right)} {9216\,{\pi }^4\,\left( Q^2 + 2\,{s_0} \right) \,
    {{{\lambda }_N}}^2} \nonumber \\ [.2cm]
 &+& \frac{ 
    3\,\ln (\frac{Q^2}{Q^2 + 2\,{s_0}})\,\left(Q^2 + 2\,{s_0} \right) \,
\left( 32\,Q^6 + 67\,Q^4\,{s_0} + 7\,Q^2\,{{s_0}}^2 - 21\,{{s_0}}^3
 \right) }
{9216\,{\pi }^4\,\left( Q^2 + 2\,{s_0} \right) \,
    {{{\lambda }_N}}^2}  \; ,
\end{eqnarray}

where one can recognize the standard logarithmic singularity arising from 
the chiral limit. In order to obtain the asymptotic behaviour of $F_1(Q^2)$ 
it is essential to expand this logarithm. In fact, there is an exact cancellation
between several terms in Eq. (16) such that the leading asymptotic term is
\begin{equation}
   \lim_{Q^2 \rightarrow \infty}
        Q^4 \; F_1(Q^2) = \frac{11 \, s_0^5}{2560 \, \pi^4  \,
        \lambda_N^2} \; ,
\end{equation}        
Qualitatively, this asymptotic behaviour agrees with
expectations.\\

There are two leading power corrections with no gluon exchange in the OPE of
the correlator Eq.(1). The one proportional to the quark condensate does not
contribute to $F_1(q^2)$, while the other, proportional to the four-quark condensate, 
leads to
\begin{equation}
\Pi^\mu(p^2,p'^2,Q^2)= \frac{8}{9} \frac{\langle \bar u u\rangle^2}{Q^2} (\frac{1}{p^2} +
\frac{1}{p'^2})  \lh{p'} \gamma^\mu \lh{p} \;+ ... \;,
\end{equation}
where $\langle \bar u u\rangle = \langle \bar d d\rangle$ has been assumed.
The double discontinuity of this term in the (s,s') complex plane vanishes, so that
it does not contribute to Eq. (14).\\

We now turn to the extraction of $F_2(q^2)$, and consider the leading order 
non-perturbative power correction to the OPE, in this case given by the 
quark condensate. 
It turns out that the contribution involving
the up-quark condensate vanishes (on account of vanishing traces),
leaving only the piece proportional to 
$\langle \bar{d} d \rangle$. The three-point function (1) becomes (see 
Fig. 3)
\begin{eqnarray}
&&{\Pi_{\langle \bar q q\rangle}}^\mu(p^2,{p'}^2,Q^2)= i \frac { 
{\langle \bar d d\rangle} }
{3 (2 \pi)^4} \left[  4 \,\int \!
\mbox{d}^4 k \frac {\mbox{Tr} \left[\lh k \gamma^\mu (\lh k-\lh q )
\gamma_\nu (\lh k- \lh p')\gamma_\alpha \right] } {(k-q)^2 (k-p')^2 k^2}\,
 \gamma^
\alpha\gamma^\nu \right. \nonumber \\ [.2cm]
&-& \left. \int \!\mbox{d}^4 {k}
\frac{\mbox{Tr}\left[\lh {k}\gamma_\nu (\lh {k}-\lh { p'})
\gamma_\alpha\right]}{k^2 (k-p')^2 q^2} \left(\gamma^\alpha\gamma^\mu
\lh q \gamma^\nu \right)
+\int\!\mbox{d}^4
 k
\frac{\mbox{Tr}\left[\lh k \gamma_\nu (\lh k-\lh p)\gamma_\alpha\right]}
{k^2 (k-p)^2 q^2} (\gamma^\alpha \lh q \gamma^\mu \gamma ^\nu)\right] \; .
\end{eqnarray}

Our choice of Lorentz structure in this case is $\lh{q} \gamma^\mu$, 
which appears in Eq.(19), as well as in Eq.(8) where it multiplies  
$F_2(q^2)$, but not $F_1(q^2)$. In fact, after some algebra
\begin{equation}
\mbox{Im} {\Pi_{\langle \bar d d\rangle}}^\mu(s,s',Q^2)
\Big|_{\mbox{QCD}}
= -\frac {\langle \bar d d\rangle} {3}\left\{
 \frac{ Q^2 s' \left( Q^2 + 3 s + s'\right) }
  {\left[ Q^4 + \left( s - s' \right)^2 + 2Q^2\left( s + s' \right)
 \right]^{\frac{3}{2}}
       } 
+\frac{1}{(2 \pi)} \frac {(s' - s)} {Q^2}
\right\} \lh q \gamma^\mu \;+\ldots \; ,
\end{equation}
and
\begin{equation}
\mbox{Im} \Pi^\mu(s,s',Q^2)\Big|_{\mbox{HAD}}=F_2(Q^2)\,
\frac{\kappa_p}{2} \left(\frac{s'} {M_N}+
M_N \right)\lh q \gamma^\mu+\ldots
\end{equation}

After substituting the above two spectral functions in the FESR Eq. (15), 
and performing the integrations one obtains
\begin{equation}
F_2(Q^2)=-\frac {\langle \bar d d\rangle} { 24 \kappa_p \, M_N\, \pi^2\,
 \lambda_N^2}
\left[2 s_0 \left( Q^2 + s_0 \right)  +
    Q^2 \left( Q^2 + 2 s_0 \right) \,
     \ln (\frac{Q^2}{Q^2 + 2 s_0})\right] \; .
\end{equation}

After expanding the logarithm there are exact cancellations between
various terms above, leaving the asymptotic behaviour
\begin{equation}
   \lim_{Q^2 \rightarrow \infty}
         F_2(Q^2) = -  \frac{\langle \bar d d\rangle} { 18 \kappa_p \, M_N\, \pi^2\,
 \lambda_N^2} (\frac{s_0^3}{Q^2} - \frac{s_0^4}{Q^4} +...)
         \; ,
\end{equation}        

Qualitatively, this asymptotic behaviour does not agree
with expectations. In fact, one expects $F_2(Q^2)$ to
fall faster than $F_1(Q^2)$ at least by a factor of
$1/Q$ \cite{JLABF2}. Quantitatively, there is also a  disagreement with data
even at intermediate  values of $Q^2$, as discussed below.\\

The results for the form factors $F_{1,2}(q^2)$, Eqs. (16) and (22), involve
the free parameters $\lambda_N$ and $s_0$. From QCD sum rules for two-point
functions involving the nucleon current (2) it has been found \cite{QCDSR}-
\cite{MNDL}
that $\lambda_N \simeq (1 - 3)\times 10^{-2}\; \mbox{GeV} ^3$, and $\sqrt{s_0}
\simeq (1.1 - 1.5)\, \mbox{GeV}$. The higher values of
$\lambda_N$ and $s_0$ come from Laplace sum rules \cite{MN}, and the lower
values are from a FESR analysis \cite{MNDL} which yields  the relation 
$s_0^3 = 192 \pi^4 \lambda_N^2$. After fitting Eq.(16) to the 
experimental data, as corrected in \cite{EXP2}, we find 
$\lambda_N = 0.011\; \mbox{GeV}^3$, and $s_0 = 1.2 \; \mbox{GeV}^2$, in 
line with the values discussed above.
Numerically, $s_0$ is well below the Roper
resonance peak, thus justifying the model used for the hadronic spectral
function, Eq.(8). The predicted form factor $F_1(q^2)$ is shown
in Fig.4 (solid line) together with the data, the agreement being quite good. A comparison of $F_2(q^2)$ from Eq.(22) with data shows a disagreement at the level of a factor two. This cannot be improved by attempting changes in the values of the free parameters $\lambda_N$ and $s_0$, and is basically a consequence of the soft $q^2$- dependence of $F_2(q^2)$, as evidenced by Eq.(23).\\

Considering now the neutron form factors, one needs to make the change
$u \leftrightarrow  d$ in Eq.(2). The perturbative QCD spectral function,
Eq.(10), involves now the combination $(\Omega_3 - \Omega_2)$. After
using the FESR Eq.(15) it turns out that $F_1(Q^2)$ for the neutron is
numerically very small and consistent with zero, except near
$Q^2=0$ where it diverges in the chiral limit. The explicit expression is
\begin{eqnarray}
F_{1n}(Q^2)&=&\frac{1}{9216\,{\pi }^4\,\left( Q^2 + 2\,{s_0} \right) \,
    {{{\lambda }_N}}^2}\,
[2\,{s_0}\,\left( - 75\,Q^6 - 207\,Q^4\,{s_0} - 106\,Q^2\,
{{s_0}}^2 +
32\,{{s_0}}^3 \right)
     \nonumber \\ [.2cm]
 &-& 
    3\,\ln (\frac{Q^2}{Q^2 + 2\,{s_0}})\,\left(Q^2 + 2\,{s_0} \right) \,
\left( 25\,Q^6 + 44\,Q^4\,{s_0} + 8\,Q^2\,{{s_0}}^2 - 6\,{{s_0}}^3
 \right) ] \; .
\end{eqnarray}

This smallness of the neutron Dirac  form factor provides a nice self-consistency
check of the method. Using $F_{1n}(Q^2) \simeq 0$, the Sachs form factors are
then proportional to $F_{2n}(Q^2)$, which is given by
\begin{equation}
F_{2n}(Q^2) = \frac {1}{48 \kappa_n M_n \pi^2 \lambda_N^2} \langle \bar u u\rangle
[ 2 s_0 (Q^2+s_0) + Q^2 (Q^2 + 2 s_0) \ln (\frac{Q^2}{Q^2+ 2 s_0})].
\end{equation}
In Fig. 5 we show the result for the electric Sachs form factor of the neutron,
together with data at low $Q^2$ \cite{EXPN}. At higher momentum transfers, there will be
a serious disagreement with experiment on account of the soft $1/Q^2$
behaviour of $F_{2n}(Q^2)$, Eq.(25). Since $G_M(Q^2)$ for the neutron appears
well fitted by the dipole formula, our QCD sum rule results do not agree
with the data. This disagreement, though, is within a factor of two, i.e. not
different from other recent QCD sum rule results \cite{BRAUN}.\\

In summary, Finite Energy QCD sum rules of leading dimensionality in the
OPE lead to Dirac form factors in very good agreement with experiment for
both the proton and the neutron. However, this is not the case for the Pauli
form factor, which exhibits a soft $Q^2$ dependence proportional to the
quark condensate. This is a welcome feature in several mesonic form factors 
where the quark condensate
contributes with a $1/Q^2$ behaviour, as expected from
experiment. Unfortunately,
this is not the case for the nucleon.  
While the results for $F_2(Q^2)$ are dissapointing, they are not worse than those from other
QCD sum rule approaches. In fact, the disagreement with data is within a factor
two. The present method at least allows to  identify clearly 
the source of discrepancy with experiment.

We comment, in closing, on the next-to-leading order (NLO) contributions
to the three-point function, Eq.(1), which were not considered here. On the
perturbative sector we expect the gluonic corrections to be small, on account
of the extra loop involved, plus the overall factor of $\alpha_s$. The NLO
power correction in the Operator Product Expansion involves the gluon
condensate. This contribution is also expected to be small, as it contains
one more loop with respect to the leading quark condensate term. In addition,
further  suppression of about one order of magnitude would arise from
numerical factors involved in the contraction
of the gluon field tensors. On the hadronic sector, the standard
single-particle pole plus continuum model adopted for the spectral function
is well justified a posteriori from the resulting value of the continuum
threshold $s_0$, well below the Roper resonance.\\

\begin{center}
{\bf Figure Captions}
\end{center}
Figure 1. The three-point function, Eq. (1), to leading order
in perturbative QCD.\\
Figure 2. Triangular and rectangular integration regions of the
Finite Energy Sum Rules, Eq. (15).\\
Figure 3. Non-vanishing terms proportional to the down-quark
condensate, Eq. (19).\\
Figure 4. Corrected experimental data on $F_1(Q^2)$
for the proton, \cite{EXP2},
together with the theoretical result from Eq.(16) (solid line). \\
Figure 5. Experimental data on $G_{E}(Q^2)$ for the neutron \cite{EXPN},
together with the theoretical results from Eqs.(24)-(25).  

\newpage
\begin{figure}[tp]
\begin{center}
\epsffile{FIG1.EPS}
\caption{}
\end{center} 
\end{figure}
\newpage
\begin{figure}[tp]
\begin{center}
\epsffile{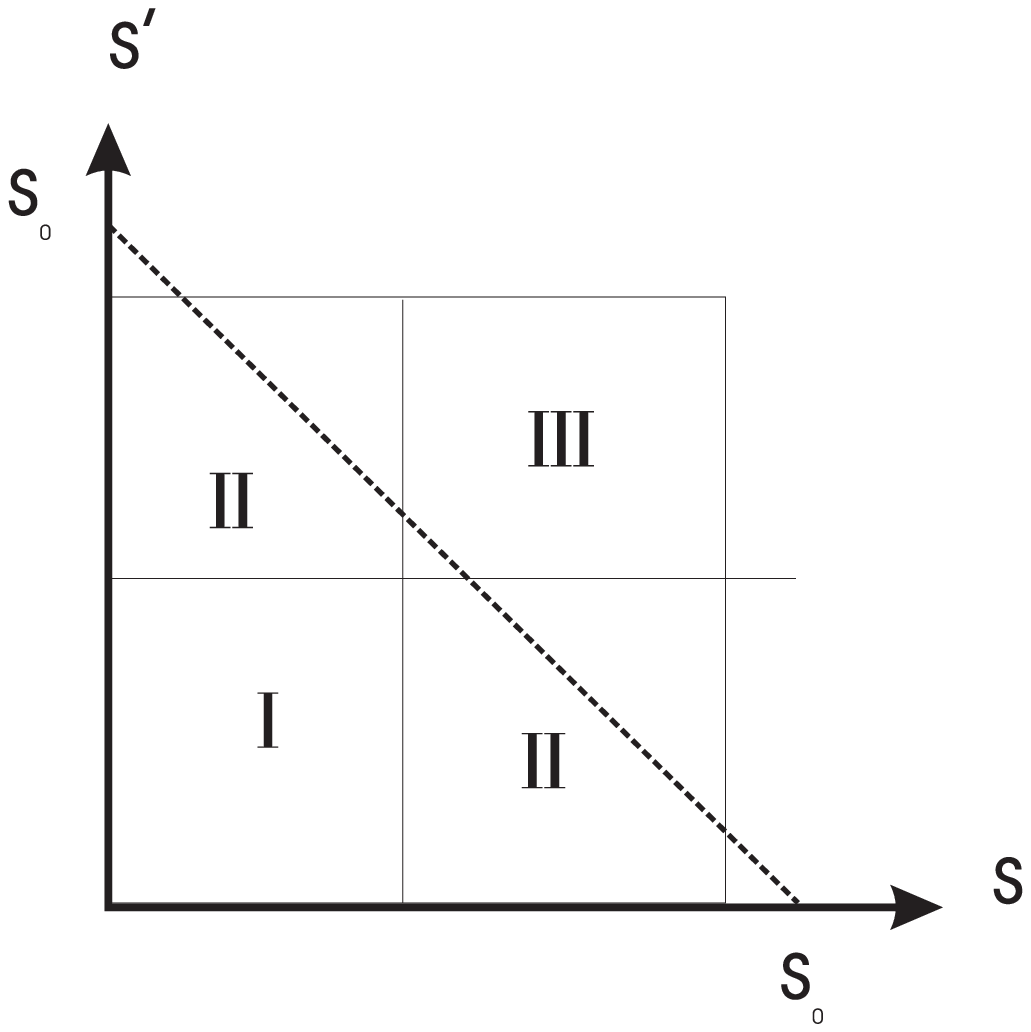}
\caption{}
\end{center} 
\end{figure}
\newpage
\begin{figure}[tp]
\begin{center}
\epsffile{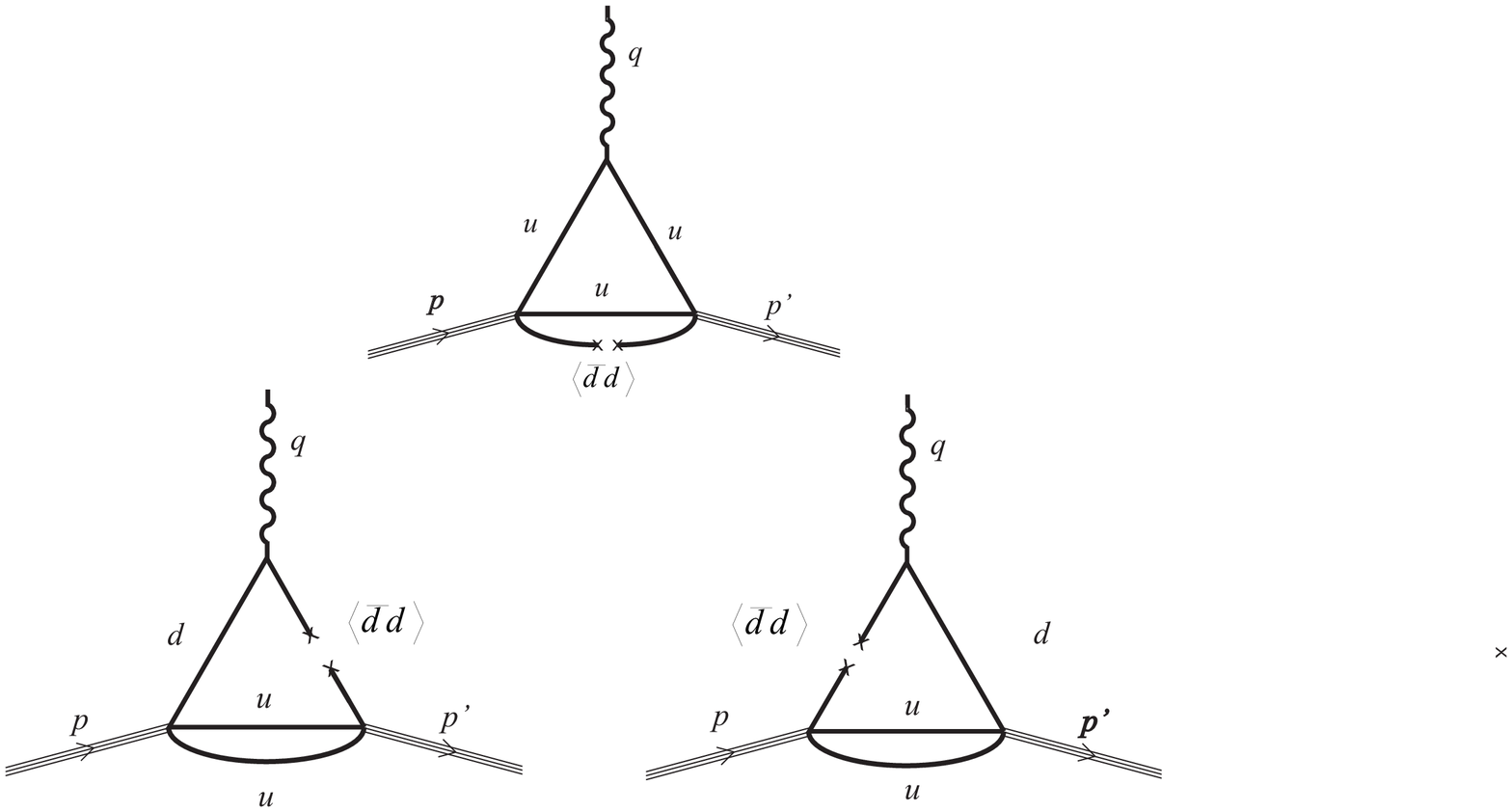}
\caption{}
\end{center}
\end{figure}
\newpage
\begin{figure}[tp]
\begin{center}
\epsffile{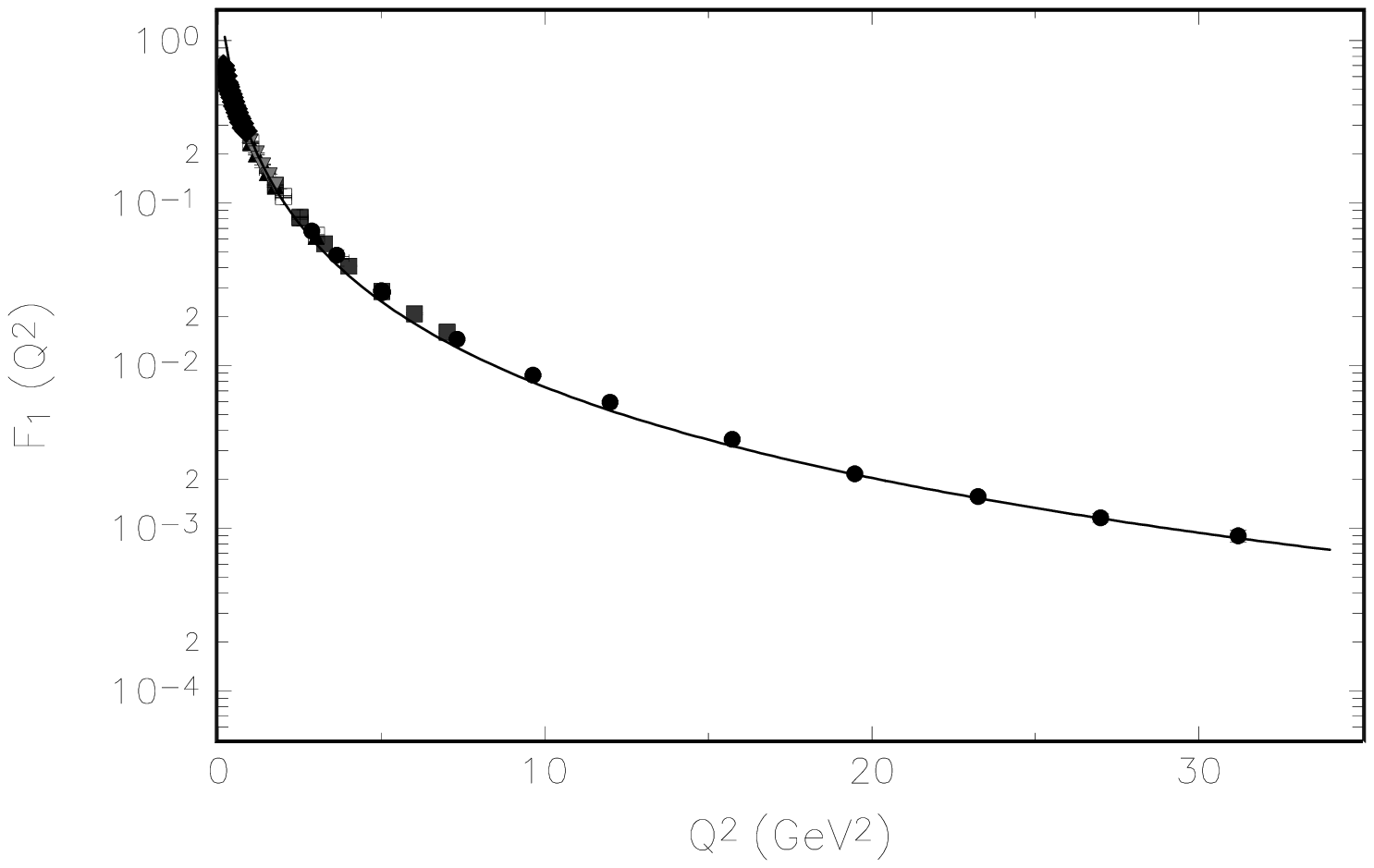}
\caption{}
\end{center}
\end{figure}
\newpage
\begin{figure}[tp]
\begin{center}
\epsffile{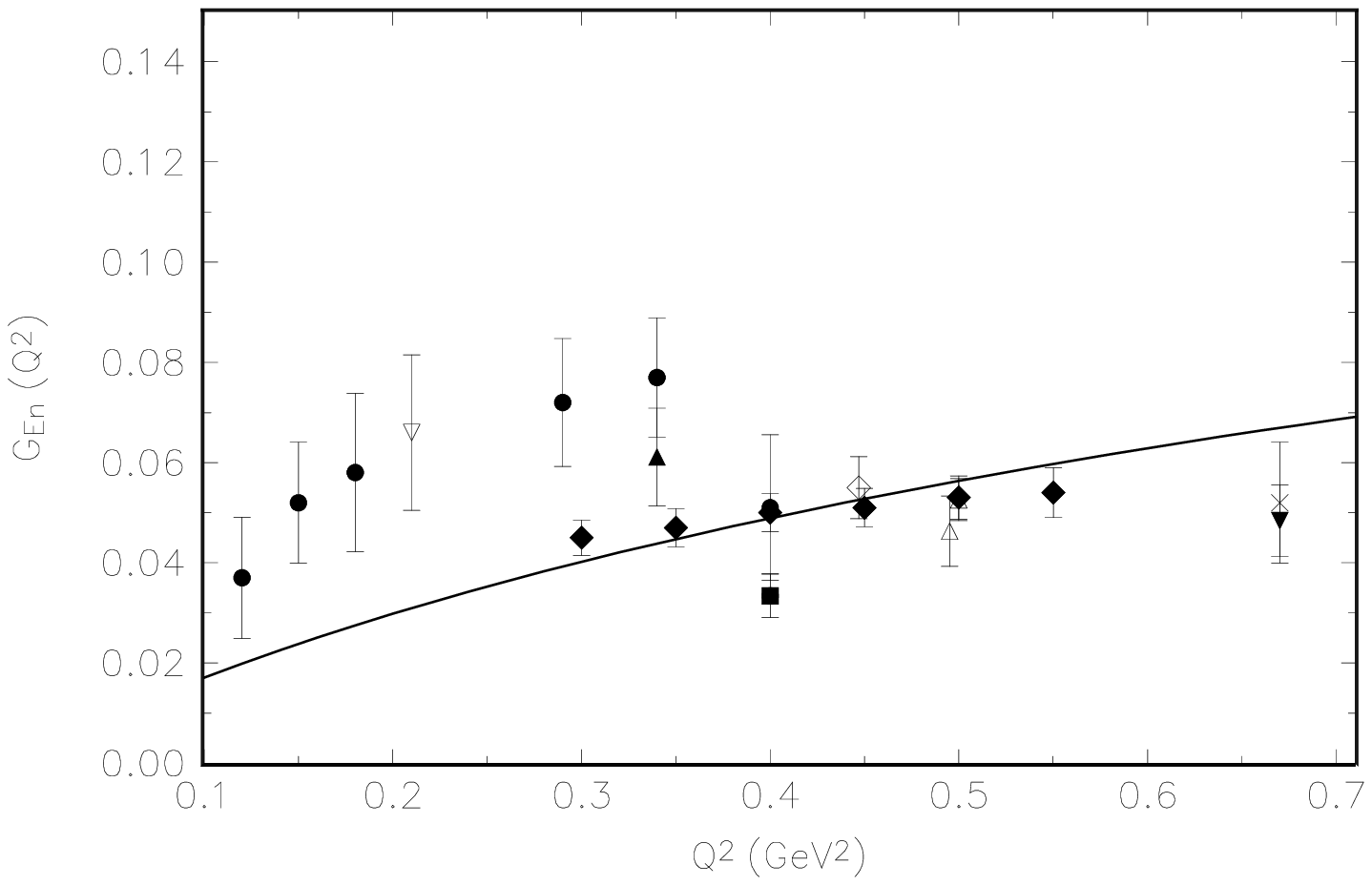}
\caption{}
\end{center}
\end{figure}

\end{document}